\input harvmac.tex

%
\def\tilde{\widetilde}
\def\bar{\overline}

\def\*{\star}
\def\({\left(}          
\def\){\right)}         
\def\[{\left[}          
\def\]{\right]}

%
%
\def\frac#1#2{{#1 \over #2}}

\def\half{{1 \over 2}}
\def\d{\partial}

\def\vev#1{\langle #1 \rangle}

\def\2pi{\hbox{$2\pi i$}}

\def\dsl{\raise.15ex\hbox{/}\kern-.57em\partial}
\def\Dsl{\,\raise.15ex\hbox{/}\mkern-.13.5mu D}

%
%
\def\th{\theta}         
         
\def\be{\beta}
\def\al{\alpha}

\def\la{\lambda}        
\def\de{\delta}         \def\De{\Delta}
\def\om{\omega}         
\def\sig{\sigma}        

%
%

       \def\CH{{\cal H}}       
\def\CJ{{\cal J}}

\def\CS{{\cal S}}

\def\si{{\rm sin}}

\def\ct{{\rm ctg}}
\def\sdif{\si \(\frac{\th_i-\th_j}{2}\)}
\def\ssum{\si \(\frac{\th_i+\th_j}{2}\)}
\def\cdif{\ct \(\frac{\th_i-\th_j}{2}\)}
\def\csum{\ct \(\frac{\th_i+\th_j}{2}\)}
\def\s1{\si \frac{\th_i}{2}}
\def\s2{\si \th_i}
%
\lref\macdo{I.G. Macdonald, S\'eminaire Lotharingien, Publ. I.R.M.A.
Strasbourg, 1988 .}
\lref\stanley{R. P. Stanley, Adv. in Math., 77, (1989) 76-115.}
\lref\Macdonald{I.G.Macdonald, {\it Orthogonal polynomial associated with
root systems}, preprint (1988).}
\lref\hald{F.D.M.Haldane, Phys.Rev.Lett. {\bf 60} (1988) 635.}
\lref\calo{
F. Calogero, J. Math. Phys. {\bf 10}, 2191, (1969).}
\lref\suth{B. Sutherland, J. Math. Phys. {\bf 12} , 246 (1971);
{\bf 12} , 251 (1971).}
\lref\YB{D.Bernard, M.Gaudin, F.D.M.Haldane and V.Pasquier, J.Phys.A {
\bf 26} (1993) 5219.}
\lref\bourbaki{N.Bourbaki, {\it El\'ements de math\'ematique},
chapitres 4,5,6, Masson 1981.}
\lref\op{M.A.Olshanetsky, A.M.Perelomov, Phys.Rep. {\bf 94} (1983)
313.}
\lref\open{D.Bernard. V.Pasquier, D.Serban, Europhys. Lett.  (1995)}
\lref\Gaudin{M.Gaudin, Saclay preprint SPhT/92-158.}
\lref\szego{G.Szeg\"o,{\it Orthogonal Polynomials}, American Mathematical
Society Colloquim Publications, Vol.XXIII}
\lref\ohta{M.Kojima, N.Ohta {\it Exact Solutions of Generalized 
Calogero-Sutherland Models}, Preprint OU-HET 239, hep-th/9603070.}
\lref\sklyanin{E.K.Sklyanin, J.Phys. A Math.Gen. {\bf 21} (1988) 2375.}
\lref\Weyl{H.Weyl, {\it The Classical Groups}, Princeton University Press, 
1946.}
\lref\Lassallea{M.Lassalle, C.R.Acad.Sci.Paris, {\bf 312} S\'erie I, 
(1991) 425.}
\lref\Lassalleb{M.Lassalle, C.R.Acad.Sci.Paris, {\bf 310} S\'erie I, 
(1990) 253.}
\lref\BGHP{D.Bernard, M.Gaudin, F.D.M.Haldane, V.Pasquier, J.Phys.A Math.Gen.
{\bf 26} (1993) 5219.}


\Title{SPhT/96/107}{
\vbox{\centerline{Some Properties of the Calogero-Sutherland Model} 
\centerline{with Reflections}}}

\centerline{D. Serban
\footnote{$^\dagger$}{Present address: Institut f\"ur Theoretische Physik,
Universit\"at zu K\"oln, Z\"ulpicher str. 77, D-50937, K\"oln, Germany.}
}

\centerline{Service de Physique Th\'eorique
\footnote{$^\ddagger$}{Laboratoire de la Direction des Sciences de la
 Mati\`ere du 
Commissariat \`a l'\'Energie Atomique.},}
\centerline{CE Saclay, 91191 Gif-sur-Yvette, France}

\vskip 1.5 truecm
Abstract

We prove that the Calogero-Sutherland Model with reflections 
(the $BC_N$ model) possesses a property of duality relating
the eigenfunctions of two Hamiltonians with different
coupling constants. We obtain a generating function
for their polynomial eigenfunctions,
the generalized Jacobi polynomials. The symmetry of the wave-functions 
for certain particular cases (associated to the root
systems of the classical Lie groups $B_N$, $C_N$ and $D_N$)
is also discussed.   

\vskip 4 truecm

Date 01/97

\vskip .5 truecm

\newsec{Introduction}

During the last years, a lot of work was devoted to the study of
the Calogero-Sutherland Hamiltonian, due to its relation to fractional
statistics in one dimension,
to the random matrix theory, etc.
The model originally proposed by Sutherland \suth\ 
describe
particles moving on a circle, with interaction proportional to the 
inverse square of the chord distance and with periodic boundary conditions.
This Hamiltonian will be referred as the periodic 
Calogero-Sutherland Hamiltonian.
This model is exactly solvable and its wave-functions, 
the Jack polynomials, were extensively studied \macdo \stanley .

Recently, a new family of models of the Calogero-Sutherland
type was proposed \op ,
describing particles on a semi-circle, interacting with one
another and with the boundaries. We call them 
Calogero-Sutherland models with reflections; several types are associated 
to several types of root systems of classical Lie algebras. 
These models are also exactly
solvable. Some properties of their polynomial eigenfunction, the Macdonald
polynomials were studied in \Macdonald .
The spectrum and the eigenfunctions 
of these Hamiltonians were used in \open\ in order to 
obtain the exact solution of a class of long-range interacting
spin chains with boundaries. 
 
These models are remarkably 
similar to the periodic one, but there are additional complications
related the loss of the translational invariance. 
One of the key properties of the periodic model is the duality
which permits to relate the wave functions corresponding to 
two different values of the coupling constant \stanley \macdo \Gaudin .
In this paper, we prove the existence of a similar duality property
for the Calogero-Sutherland Hamiltonian with reflections.
The initial motivation of this work was in obtaining the correlation 
functions of the Calogero-Sutherland models with reflections.

The plan of the paper is the following: the next section is devoted to
the presentation of the model, in section 3 we present a method for deriving
the polynomial eigenfunctions and in the section 4 we emphasize 
the connection between the eigenfunctions of these Hamiltonians and the Jacobi functions. The section 5 is devoted to the 
proof of the duality property. In section 6 we use this 
property in order to derive an expansion formula for the kernel
which intertwines between the two dual models. 
  
\newsec{The Model}
 
Following the references  \op\ \Macdonald, a Hamiltonian of
Calogero-Sutherland type can be defined for each
root system of a classical Lie algebra.
Let $V$ denote a $N$ dimensional vector space with an orthonormal 
basis $\{e_1,...
,e_N\}$ and let $R=\{\al \}$ be a root system in $V$, 
with $R_+$ the set of positive roots. Let $\th$ denote the
vector $(\th_1,...,\th_N)$ and $\th \cdot \al$ its scalar product with 
the vector $\al$. The generalized CS Hamiltonian is~:
\eqn\CS{
H=-\sum_{i=1}^N \frac{\d^2}{\d \th_i^2}+
\sum_{\al\in R_+} \frac {g_{\al}}{{\rm sin}^2 (\th\cdot\al/2)}\;,
}
where $g_{\al}$ is constant on each orbit of the Weyl group {\it i.e.}
it has the
same value for the roots of the same length.

The periodic model \suth\ correspond to the 
root system of type $A_{N-1}$.
It describes interacting particles on a circle, with the
positions specified by angles $\th_i$ ranging from $0$ to $2\pi$ and with 
periodic boundary conditions.

The reflection models are associated to the
four infinite series of root systems
$D_N$, $B_N$, $C_N$ and $BC_N$.
A list of the main characteristics of these series of 
root systems is given in the appendix A1. 
  
The most general Hamiltonian is the one associated to the 
{\it non-reduced}(i.e. that includes roots 
which are proportional, as $\al$ and $2\al$) $BC_N$ root system; 
the others can be obtained from it
by setting the coupling constants to some special values.

The $BC_N$ Hamiltonian describes $N$
particles on a semi-circle, with positions specified by the angles 
$0\leq\theta_1,...,\theta_N
\leq\pi$~:
\eqn\ham{\eqalign{
H=&-\sum_{i=1}^N \frac{\d^2} {\d \th_i^2}+\be(\be-1)\sum_{i\neq j=1}^N
\[\si^{-2}\(\frac{\th_i-\th_j}{2}\)+\si^{-2}\(\frac{\th_i
+\th_j}{2}\)\] \cr
& +\sum_{i=1}^N \[c_1(2c_2+c_1-1)\si^{-2}\frac{\th_i}{2}
+4c_2(c_2-1)\si^{-2} \th_i \]\;. \cr
} }
This Hamiltonian has three independent coupling constants $\be,c_1,c_2$, 
corresponding to the roots of length $2$, $1$ and $4$ respectively. 

Compared to the periodic version, the potential part of this
Hamiltonian contain
a new two-body term corresponding to the interaction between the particle
$i$ and the "image" of the particle $j$ through the
reflection $\th_j\rightarrow -\th_j$.
Using the relation $\sin^{-2}x+\cos^{-2}x =4\sin^{-2}(2x)$,
the one-body part of the potential can be separated into 
couplings of the particles to the two boundaries $\th=0,\pi$, 
with independent coupling constants related to $c_1,c_2$.

The other cases are obtained by setting to zero one (or both) of the coupling
constants $c_1$, $c_2$~:
\eqn\difcase{
\matrix{B_N\;: \quad & c_2=0 \cr
         C_N\;: \quad & c_1=0 \cr
         D_N\;: \quad & c_1=c_2=0 \cr}
}
The symmetry to be
imposed to the wave functions depend on the root system we consider.

The ground state wave function of this Hamiltonian is \open ~:
\eqn\ground{
\De(\th)=\prod_{i=1}^N \[\si^{c_1} \frac{\th_i}{2}~ \si^{c_2} \th_i\]
\prod_{i<j} \[\si^\be \(\frac{\th_i-\th_j}{2}\)~
\si^\be\(\frac{\th_i+\th_j}{2}\)\]\;.}
Remark that this ground state is well defined for $\be,c_1,c_2>-1/2$.

It is convenient to define a gauge transformed
Hamiltonian
by $\CH =\De(\th)^{-1}
H\De(\th)-E_0$, with $E_0=\sum_{i=1}^N (\be(N-i)+c_1/2+c_2)^2$ the 
ground state energy of $H$. We obtain~:
\eqn\gauge{
\CH=-\sum_{i=1}^N \d_i^2-\be \sum_{i\neq j}\[\cdif+\csum\]\d_i-
\sum_{i=1}^N\[c_1\ \ct \frac{\th_i}{2}+2 c_2\ \ct \th_i\]\d_i\;.
}
Let us mention that higher order conserved quantities exist for this model.
They are generated by the quantum determinant of the monodromy matrix 
obeying the reflection equation \sklyanin . Their construction parallels the
one of the conserved quantities of the periodic model \BGHP .
The monodromy matrix for the spin chains associated to this model was 
constructed in \open .

\newsec{Symmetry of the Eigenstates of $\CH$}

In this section, we present basis of polynomials
in the variables $z_j^{\pm1/2}=e^{\pm i\th_j/2}$ in which the  
Hamiltonian $\CH$ is triangular \Macdonald . 
This basis can serve for determining the
eigenvalues and to find the eigenfunctions.   

We emphasize that different symmetries can be assigned to these eigenfunctions,
depending on the values of the coupling constants $c_1,\ c_2$.
These symmetries can be best understood in terms of root systems  \Macdonald ,
as invariances under
transformations defined by the Weyl
group. The wave functions are naturally indexed by the dominant
weights of the root systems $BC_N$ (or $D_N$, $B_N$, $C_N$ for the particular
values of the coupling constants mentioned in \difcase ).

We start this section by a brief review (for more details see for example 
\bourbaki ) 
of some of the notions related to the root systems.

Let $V$ be a $N$ dimensional vector space with an orthonormal 
basis $\{e_1,...
,e_N\}$ and $\al$ a root system in $V$. Let $X$ be the reunion of hyperplanes
orthogonal to one of the roots $\al$. A chamber is a connected component
of $V-X$. 
Let $(\al_1,...,\al_l)$ be the basis corresponding
to a chamber $C$ ($(\al_i,x)>0$ for $x\in  C$)
and $\al_i^V=2\al_i/(\al_i,\al_i)$.
The vectors $\bar \om_i$ with the property $(\al_i^V,\bar \om_j)=\de_{ij}$
are called the fundamental weights. The dominant weights are defined as being
linear combinations of the fundamental weights with non-negative integer
coefficients, $\la=\sum_{i=1}^l k_i\bar \om_i$. When $l=N$, as in the cases 
considered here, the 
dominant weights can 
equally be characterized by the set of coordinates $\la_1,...,\la_N$
of $\la$ with respect to the the orthogonal system $e_1,...,e_N$; 
$\la_i=(\la,e_i)$.

We remind that the Weyl group is the group generated by the reflections
with respect to the hyperplanes orthogonal to the roots. 

The the main characteristics of the 
root systems we consider, as well as the allowed values of $\la_1,...,\la_N$,
are presented in appendix A1. 

A {\it partial} ordering can be defined for the dominant weights.
$\la>\mu$ if $\la $, $\mu$ are dominant weights  
and $\mu=\la-\al$ with  $\al$ a positive root.

Consider now the Hamiltonian $\CH$ written in the variables $z=e^{i\th}$~:
\eqn\ggauge{
\CH=\sum_{i=1}^N (z_i \d_{z_i})^2 +\beta \sum_{i\neq j}
(w_{ij}+\bar w_{ij})z_i \d_{z_i} + \sum_{i=1}^N (c_1w_{i0}+2c_2\bar w_{ii})
z_i\d_{z_i}\;,
}
with $w_{ij}=(z_i+z_j)/(z_i-z_j)$, $\bar w_{ij}=(z_i+z_j^{-1})/
(z_i-z_j^{-1})$ and  $w_{i0}=(z_i+1)/(z_i-1)$.
It is possible to construct eigenfunctions of $\CH$, polynomial
in the variables $z_j^{\pm1/2}=e^{\pm i\th_j/2}$ and symmetric
under the transformations defined by the Weyl group. 
To achieve that, start from the monomials~:
$$z_1^{\la_1}...z_N^{\la_N}$$
and  define the symmetrized monomials~: 
\eqn\symmon{m_\la=\sum_{s\in W} z_1^{s(\la_1)}...z_N^{s(\la_N)}\;,}
where the sum is over the elements of the Weyl group $W$,
each distinct monomial occurring only once,
and $\la$ denote a dominant weight of the root system under consideration.
The Hamiltonian $\CH$ is triangular on the basis of 
symmetrized monomials $m_\la$~:
$$\CH m_\la=E_\la m_\la +\sum_{\mu<\la}c_{\la,\mu} m_\mu. $$
One can check that there are no poles generated by
the $w_{ij}$ factors. The symmetry of $m_\la$ insures that these poles
disappear and
lower order monomials are generated. The typical example is~:
$$\frac{z_1+z_2}{z_1-z_2}(z_1^{n_1}z_2^{n_2}-z_1^{n_2}z_2^{n_1})
=z_1^{n_1}z_2^{n_2}+2z_1^{n_1-1}z_2^{n_2+1}+...+z_1^{n_2}z_2^{n_1},$$
where $n_1-n_2$ is a positive integer. The terms containing
$w_{ij}$, $\bar w_{ij}$,
$w_{i0}$ and $\bar w_{ii}$ generate lower order symmetric monomial
of the type $m_{\la -\al}$ with $\al$ equal to $e_i-e_j$, $e_i+e_j$,
$e_i$ and $2e_i$ respectively.

As $\CH$ is triangular, the eigenvalues $E_\la$ are easily derived~:
\eqn\energ{
E_{\la}=\sum_{i=1}^N \la_i(\la_i+2\be(N-i)+c_1+2c_2).
}
The restrictions on the values of the momenta $\la_i$, coming
from the fact that $\la$ is a dominant weight, are discussed
in the two appendices.
 
\newsec{Jacobi Polynomials}

The polynomial eigenfunction of $\CH$ with $BC_N$ symmetry 
considered in the previous are also symmetric polynomials in the variables 
$x_j=\cos\th_j$. They are multivariate generalizations of the 
Jacobi polynomials \Lassallea .

Let us start with the simplest cases $\be=0$ or $1$, when 
the Hamiltonian \ham\ decouples to a 
sum of one-particle
terms $H_1$. After a gauge transform $\varphi(\th) H_1 \varphi^{-1}(\th)$,
with $\varphi(\th)=\si^{c_1} \frac{\th}{2}~ \si^{c_2} \th$, 
we obtain the one-particle Hamiltonian~:
\eqn\unipart{
\CH_1=-\frac{d^2}{d\th^2}-
(c_1\ct \frac{\th}{2}+2c_2 \ct \th)\frac{d}{d\th}.
} 
The eigenfunctions of $\CH_1$ satisfy the hypergeometric differential
equation in the variable $x=\cos \th$~:
\eqn\Jacobi{
(1-x^2)\frac{d^2y}{dx^2}-[c_1+(c_1+2c_2+1)x]\frac{dy}{dx}
+n(n+c_1+2c_2)y=0.
}
The Jacobi polynomials $P_n^{(a,b)}(x)$, with~:
\eqn\abc{a=c_1+c_2-1/2, \quad
b=c_2-1/2} 
and $n$ non-negative integer, are solutions of this equation.
We will use $a,b$ to index the wave functions and continue to use 
$c_1,c_2$ as coupling constants in the Hamiltonian.
 
The Jacobi polynomials form a basis for the functions defined on the 
interval $[-1,1]$, orthogonal with respect to the scalar product~:
\eqn\scap{
\vev{f(x),g(x)}=\int_{-1}^1 dx\ (1-x)^{a-b}(1-x^2)^b f(x)g(x)\;.
}
The second (non-polynomial) solution of \Jacobi\ is
given by the Jacobi's function of second kind, $Q_n^{a,b}(x)$. 
For a detailed description of the Jacobi polynomials and Jacobi functions 
see ref. \szego .
We retain the following expansion property~:
\eqn\expja{
\sum_{n=0}^\infty \{h_n^{(a,b)}\}^{-1}P_n^{(a,b)}(x)Q_n^{(a,b)}(y)=
\half\frac{(y-1)^{-a+b}(y^2-1)^{-b}}{y-x}
}
where $h_n^{(a,b)}$ is the norm of the Jacobi 
polynomials with respect to the scalar product \scap\ 
and $x=\cos\th$ and $y=\cos\phi$.

The Jacobi polynomials play the same role for the $BC_N$ model as the power 
functions $z^n$ for the periodic model. In particular, bosonic (fermionic)
wave functions can be obtained by symmetrisation (antisymmetrisation)
of products of Jacobi polynomials~: 
\eqn\boso{
\CJ_{\la_1,...\la_N}^{(a,b)}(\cos\th_1,...,\cos\th_N;0)=
d_\la (0,a,b)\sum_{\sig\in S_N} P_{\la_1}^{(a,b)}(\cos\th_{\sig_1})...
P_{\la_N}^{(a,b)}(\cos\th_{\sig_N})
}
and respectively~:
\eqn\fermi{
\CJ_{\la_1,...\la_N}^{(a,b)}(\cos\th_1,...,\cos\th_N;1)=
d_\la(1,a,b) \frac{
{\rm det} \(P_{\la_i+N-i}^{(a,b)}(\cos \th_j)\)}
{\prod_{i<j} \sdif \ssum }\;,
}
where $d_\la (\be,a,b)$ are normalization constants to be fixed later.

The functions defined by the relation \fermi\ are analogue to the 
Schur polynomials. In the particular case $c_1=0,\ c_2=1$ (or $a=b=1/2$)
they are, up to a normalization constant, the characters of the 
symplectic group \Weyl .

For a generic value of $\beta$, M.Lassalle \Lassallea\ showed that 
there are eigenfunctions of $\CH$, uniques up to a normalization,
which have a triangular expansion on the Jack polynomials~:
\eqn\triJack{
\CJ^{(a,b)}_\la(x.;\be)=\sum_{\mu \subseteq \la}c_{\la \mu}J_\mu(x.;\be)\;,}
where $\mu \subseteq \la$ means $\mu_i \leq \la_i$ for all $i$.
They were named generalized Jacobi polynomials. 
We choose their normalization such that $c_{\la \la}=1$.
Here, we used a result of \Lassalleb\ sect. 3 to relate the Jack polynomials
in the variables $x=\cos\th$ to the ones in the variables 
$\sin^2 \th/2=(1-x)/2$ used in \Lassallea .

A method to express these polynomials in terms of Jack polynomials 
(associated to the $A_{N-1}$ root systems) was equally proposed in \ohta ,
using a bosonic representation of the Calogero-Sutherland
Hamiltonian.

\newsec{Duality}

It was proven by I.G.Macdonald \macdo\ and by M.Gaudin \Gaudin\ that the 
eigenfunctions
of the periodic model for two different coupling constants
($\be$ and $1/\beta$) are in correspondence. We show that a
similar property holds for the $BC_N$ model.  Since the method of
\macdo is  difficult to parallel in the $BC_N$ case, we employ 
the method proposed by M.Gaudin. 

Consider two set of independent variables $\th.=\{\th_i,\ i=1,N\}$ 
and $\phi.=\{\phi_m,\ m=1,M\}$ and the kernel~:
\eqn\ker{
K_{NM}(\th.;\phi.)=\prod_{m=1}^M \prod_{i=1}^N
\si \(\frac{\th_i-\phi_m}{2}\) \si \( \frac{\th_i+\phi_m}{2}\).
}
This kernel intertwines between the Hamiltonians
$\CH_N(\th.;\be,c_1,c_2)$ and $\CH_M\(\phi.;1/\be,\tilde c_1
,\tilde c_2 \)$~:
\eqn\sym{\eqalign{
-\be^{-1/2}\[\CH_N\(\th.;\be,c_1,c_2\) -\be MN(2N-1)/2-
MN\(c_1+2c_2\)\] K_{NM}(\th.;\phi.)= &      \cr
\be^{1/2} \[\CH_M
\(\phi.;1/\be ,\tilde c_1,\tilde c_2 \) 
 -\be^{-1} NM(2M-1)/2-
MN(\tilde c_1 +2 \tilde c_2) \] K_{NM}(\th.;\phi.)&\;. \cr
}}
where the dual values $\tilde c_1$ and $\tilde c_2$ are defined by~:
\eqn\dualval{
\tilde c_1=c_1/\be, \quad \tilde c_2=(2c_2-\be+1)/2\be\;.}
The proof uses repeatedly the identity  $\ct x\ \ct y+\ct y\ \ct z+
\ct z\ \ct x =1$ for angles satisfying $x+y+z=0$.

Let us take first $M=1$ and evaluate the action of 
the kinetic operator~:
$\d_\th^2=\sum_i \d_i^2$ on $K_N=K_{N1}$~:
\eqn\era{
\d_iK_{N}=\half\[\ct \(\frac{\th_i-\phi}{2}\)
+\ct \(\frac{\th_i+\phi}{2}\)\]K_{N}
}
\eqn\erb{
\d_\th^2K_N=\[-\frac{N}{2}+\half \sum_i
\ct \(\frac{\th_i-\phi}{2}\) \ct \(\frac{\th_i+\phi}{2}\)\]K_N
}
Using the property $\ct x\ \ct y+\ct y\ \ct z+
\ct z\ \ct x =1$ for the angles $x=(\th_i-\phi)/2$,
$y=-(\th_i+\phi)/2$ and $z=\phi$, we obtain~:
\eqn\cotan{
 \ct \(\frac{\th_i-\phi}{2}\) \ct \(\frac{\th_i+\phi}{2} \)K_N
 =(-2 \ct \phi \d_\phi-1) K_N }
so the kinetic term is~:
\eqn\kin{
\d_\th^2 K_N=-\(N+\ct \phi \d_\phi \)K_N.}
The one-body part of the potential in the 
variables $\th$ can be transformed into a derivative acting on
the variable $\phi$~:
\eqn\abody{
 - \sum_i \(c_1 \ct \frac{\th_i}{2} +2c_2 \ct \th_i\)\d_i K_N=   
 \[\(c_1 \ct \frac{\phi}{2}+2c_2 \ct \phi \) \d_\phi  + c_1+2c_2 
\] K_N\;.
}
The two-body part of the potential in the variables $\th$
reconstitutes the kinetic part of the Hamiltonian in the variable $\phi$~:
\eqn\bbody{\eqalign{
&-\be \sum_{i<j} \[\cdif(\d_{\th_i}-\d_{\th_j})+
 \csum(\d_{\th_i}+\d_{\th_j})\] K_N =  \cr 
& \be N(N-1)K_N -\frac{\be}{2} \sum_{i<j} \(
C_{i+} C_{j+}+
C_{i-} C_{j-}- C_{i-}C_{j+} -
C_{i+} C_{j-} 
\)  K_N = \cr
& \be N(N-1)K_N + \frac{\be}{4}\[ \( \sum_{i}\(
C_{i-}- C_{i+}\) \)^2 - \sum_i \(C_{i-}- C_{i+}
\)^2\] K_N  \cr
&  = \be\[N(N-\half)+\d^2_\phi+\half
\ct \(\frac{\th_i-\phi}{2}\) \ct \(\frac{\th_i+\phi}{2}\) \] K_N \;, \cr
}}
where $C_{i\pm}=\ct \(\frac{\th_i \pm \phi}{2}\)$.
Using \kin , \abody\ et \bbody\ we obtain~:
\eqn\htheta{\eqalign{
\CH(\th;\be, & c_1,c_2) K_N=    \cr
 =\be & \[\d_\phi^2 + 
\( \tilde c_1 \ct \frac{\phi}{2} +2\tilde c_2
 \ct \phi\) \d_\phi \] K_N
+(\be N^2+(c_1+2c_2-\be+1)N) K_N   \cr
}}
The right hand side of this equation is, up to a constant,
the Hamiltonian for one particle of coordinate $\phi$, with the 
new coupling constants $\tilde c_1=c_1/\beta$ and $\tilde c_2=
(2c_2-\be+1)/2\be$.

Let us take now $M$ variables $\phi$, $M>1$ and
$K_{NM}(\th_1,...,\th_N;\phi_1,...,\phi_M)=$
$ \prod_{i=1}^M K_N(\th_1,...,\th_N;\phi_i)$.
The potential part of the Hamiltonian 
$\CH(\th;\be, c_1,c_2)$ is a first order derivative, so its
action on the kernel $K_{NM}(\phi_1,...,\phi_M)$ is additive. The second
order derivatives in the kinetic energy operator generate
crossed terms which correspond to the two-body terms of the
Hamiltonian in the variables $\phi_1,...,\phi_M$~:
\eqn\cross{\eqalign{
\d_{\th}^2 & K_{NM}  =\sum_{i=1}^N \[-\frac{M}{2} +\frac{1}{4} 
\(\sum_{m =1}^M
\(C_{i m -} +C_{im +}\)\)^2 
 -\frac{1}{4} \sum_{m=1}^M
\(C_{im -}^2 +C_{im +}^2\) \]K_{NM}  \cr
 =& \[-\frac{NM}{2} +\half  \sum_{i,m}
\ct \frac{\th_i-\phi_m}{2} \ct \frac{\th_i+\phi_m}{2}  
 +\sum_{m\neq n}
\(\ct \frac{\phi_m-\phi_n}{2}+\ct \frac{\phi_m+\phi_n}{2}
\)\d_{\phi_m}\] K_{NM}  \cr
}}
Here, we used a calculation of the same type as in \bbody , but 
involving the index
$m$ of  $C_{im\pm}=\ct \(\frac{\th_i \pm \phi_m}{2}\)$ 
instead of the index $i$.
The last term in \cross\ is proportional to the two-body interaction in 
variables $\phi$. 

The full result is obtained by collecting the partial results in 
\bbody , \cotan\ (summed over the $M$ variables $\phi_m$) and \cross ~:
\eqn\hthphi{\eqalign{
&\CH_N(\th.;\be,c_1,c_2) K_{NM}(\th.;\phi.)=  \cr
&   \[-\be \CH_M\(\phi.;1/\be,\tilde c_1,\tilde c_2
\)+\be MN(N-1)+NM^2+MN(c_1+2c_2) \] K_{NM}(\th.;\phi.)\;. \cr 
}}
This is equivalent to the result announced in the equation \sym .
In the next section, this property will be used in order to
obtain the expansion of $K(\th.;\phi.)$ in terms of the eigenfunctions
of $\CH_N \(\th.;\be,c_1,c_2\)$ and $\CH_M
\(\phi.;1/\be,\tilde c_1,\tilde c_2\)$.

Using a similar method, we can derive another result whose periodic
analogue is well known \macdo \stanley . It concerns
the expansion of $K^{-\beta}(\th.;\phi.)$ on the eigenfunctions 
of $\CH \(\th;\be,c_1,c_2\)$~:
\eqn\dualbis{\eqalign{
&\[\CH_N\(\th.;\be,c_1,c_2\)-\CH_M\(\phi.;\be,-c_1,-c_2+\be\)\]
K_{NM}^{-\beta}(\th.;\phi.) \cr
&=-\[\be^2 MN(M-N+1)-\be MN(c_1+2c_2)\]
K_{NM}^{-\beta}(\th.;\phi.)\;. \cr
}}
We can further transform this expression, by noting the 
following property~:
\eqn\ggauge{
\psi^{-1}(\phi.) \CH_M\(\phi.;\be,c_1,c_2\) \psi(\phi.)=
\CH_M\(\phi.;\be,-c_1,-c_2+1\)-C_1\;,
}
where $\psi(\phi.)=\prod_{i=1}^M\(\sin^{-2c_1}(\phi_i/2)
\sin^{-2c_2+1}\phi_i\)$ and $C_1=M(c1+2c_2-1)\(\be(M-1)+1\)$.
This allows to rewrite \dualbis  as~:
\eqn\dualbbis{
\[\CH_N\(\th.;\be,c_1,c_2\)-\CH_M\(\phi.;\be,c_1,c_2-\be+1\)
+C_2\] \psi(\phi.) K_{NM}^{-\beta}(\th.; \phi.) =0\;,
}
where the constant $C_2=\be^2MN(M-N+1)-(c_1+2c_2)\be 
MN+M(c1+2c_2-2\be+1)\(\be(M-1)+1\)$.

\newsec{Expansion Formula for $K_{NM}(\th.;\phi.)$}

The kernel \ker\ is a polynomial in both sets of variables
$y_i=\cos\th_i$ and $w_m=\cos\phi_m$~:
\eqn\dualJack{
K_{NM}(\th.;\phi.)=\prod_{m=1}^M \prod_{i=1}^N 
\si \(\frac{\th_i-\phi_m}{2}\) \si \( \frac{\th_i+\phi_m}{2}\) =2^{-NM}
\prod_{m=1}^M \prod_{i=1}^N (y_i-w_m)\;.}
It plays the role of a generating function for the generalized 
Jacobi polynomials.
Using the equation \sym\ we prove the following property, similar to the dual
expansion of the Jack polynomials \stanley \macdo~:
\eqn\expansion{
K_{NM}(\th.;\phi.)=2^{-NM}\;\sum_{\la} (-1)^{|\tilde \la|}\; 
 \CJ^{(a,b)}_{\la}(y.;\be)\;
\CJ^{(\tilde a,\tilde b)}_{\tilde \la}(w.;1/\be)\;,}
where~:
$$a=c_1+c_2-1/2,\quad b=c_2-1/2; \qquad \tilde a=(a-\be +1)/\beta,\quad 
\tilde b=(b-\be+1)/\be $$  
and the symbol $\tilde \lambda$ denotes the partition with parts $\tilde \la_k
=N-\la'_{M-k+1}$, where $\la'$ denotes the partition conjugate to 
$\la$.

Let us proof this relation. 
Call $A_\la(w.;\be,a,b)$ the coefficients of the expansion 
of $K_{NM}(\th.;\phi.)$
on eigenfunctions of  $\CH_N\(\th.;\be,c_1,c_2\)$,
\eqn\symdual{
K_{NM}(\th.;\phi.)=2^{-NM}\;\sum_\la A_\la(w.;\be,a,b)\; 
\CJ^{(a,b)}_{\la}(y.;\be)\;.}  
It follows from the equation
\sym\ that they are eigenfunctions
of $\CH_M\(\phi.;1/\be,\tilde c_1, \tilde c_2\)$, corresponding to the same
eigenvalue as $\CJ^{(\tilde a,\tilde b)}_{\tilde \la}(w.;1/\be)$.
As the energy levels can be degenerate, we still have to prove
that these $A_\la(w.;\be,a,b)$ is proportional to 
$\CJ^{(\tilde a,\tilde b)}_{\tilde \la}(w.;1/\be)$ and to determine the
proportionality constant. To prove this, one can use the duality
property of the Jack polynomial and exploit their relation to the
generalized Jacobi polynomials \triJack . 

The expression \dualJack can be expanded on the Jack polynomials \stanley ~:
\eqn\JJack{
K_{NM}(\th.;\phi.)=2^{-NM}\;\sum_\la (-1)^{|\tilde \la|}\;
J_\la(y.;\be)\; J_{\tilde 
\la}(w.;1/\be)\;,}
where we used the relation between the Jack polynomials with arguments $w$
and with arguments $w^{-1}$, 
$$ J_{\tilde \la}(w.;1/ \be)=\prod_{m=1}^M w_m^N\ J_{\la'}(w^{-1}.;1/\be \;.)$$
This property can be easily verified using the triangularity of 
$J_\la(y;\be)$ in the
basis of symmetric monomials $m_\la$ and the fact that boths sides are 
eigenfunctions of the periodic Calogero-Sutherland corresponding to the 
same eigenvalue.

In the equation \symdual\ we can expand the generalised Jacobi polynomials 
on the Jack polynomials, to obtain~:
\eqn\dualJacobi{
K_{NM}(\th.;\phi.)=2^{-NM}\;\sum_\la A_\la(w.;\be,a,b)\sum_{\mu \subseteq \la}
 c_{\la, \mu}\; J_\mu(y.;\be)\;,
}
where $\mu \subseteq \la$ means $\mu_i\leq \la_i $ for all $i$.
From the relation \JJack\ and the orthogonalithy of Jack polynomials, we
have~:
 \eqn\relat{
 J_{\tilde \mu}\;(w.;1/\be)=(-1)^{|\mu|}\sum_{\la \subseteq \mu} c_{\la \mu} 
\; A_\la(w.;\be,a,b)\;.} 
Inverting this expansion we obtain~:
\eqn\invert{
A_\la(w.;\be,a,b)=\sum_{\tilde \mu \subseteq \tilde \la} (-1)^{|\mu|}\;
c'_{\la  \mu}\; J_{\tilde \mu}(w.;1/\be)\;,}
with $c'_{\la \la}=1$. 
As the generalized Jacobi polynomials $\CJ^{(\tilde a,\tilde b)}_{\tilde \la}
(w.;1/\be)$ are uniquely defined as being 
eigenfunctions of the hamiltonian $\CH_M\(\phi.;1/\be,\tilde c_1, 
\tilde c_2\)$ triangular in the basis of Jack polynomials and
with the coefficient $c_{\la \la}=1$, we conclude that~:
$$ A_\la(w.;\be,a,b)=(-1)^{|\tilde \la|}\; 
\CJ^{(\tilde a,\tilde b)}_{\tilde \la}
(w.;1/\be)\;,$$
which proves \expansion .
Remark that the expansion in \expansion contains 
just a finite number of terms, corresponding 
to partitions $\la$ included in the partition $M^N$.

The equations \dualbis and\dualbbis can also be related
to expansion relations for $K_{NM}^{-\be}(\th.;\phi.)$ and
$\psi(\phi.)\;K_{NM}^{-\be}(\th.;\phi.)$, involving probably generalised 
hypergeometric 
functions which are not polynomials. 
The simplest case of \dualbbis\ is $\be=N=M=1$, when the associate expansion
relation is the expansion property of the Jacobi functions
\expja .

\newsec{Acknowledgements}

I wish to thank D.Bernard, M.Gaudin, F.Lesage and V.Pasquier for many 
discussions and for reading the manuscript.  

\appendix{A1}{Root Systems}

The main characteristics of the root systems $D_N$, $B_N$, $C_N$ and $BC_N$
are the following~:

i) {\it The $D_N$ root system.} The positive roots are~: 
$$e_i-e_j,\quad e_i+e_j, \quad 1\leq i<j\leq N\;.$$
The fundamental weights are~:
$$\bar \om_i =e_1+...+e_i, \quad 1\leq i\leq N-2$$
$$\bar \om_{N-1} =\half(e_1+...+e_{N-2}+e_{N-2}-e_N)$$
$$\bar \om_{N} =\half(e_1+...+e_{N-2}+e_{N-2}+e_N)\;.$$
The dominant weights of $D_N$ are indexed by  
$\la_1\geq...\geq|\la_N|\geq0$
all integers or all half-integers. $\la_N$ can be positive or negative.
The action of the Weyl group on $\la_i$ is generated by the permutations $s_{ij}\la_i=\la_j$
and by $\bar s_{ij}\la_i=-\la_j$.

ii) {\it The $C_N$ root system} has the positive roots~:
$$e_i-e_j,\quad e_i+e_j, \quad 1\leq i<j\leq N;
\quad 2e_i \quad (1\leq i\leq N).$$
The fundamental weights are~:
$$\bar \om_i =e_1+...+e_i, \quad 1\leq i\leq N$$ 
and the dominant weights
are characterised by  
$\la_1\geq...\geq\la_N\geq0$ a set of positive (or zero) integers.
The Weyl group contains, beside the permutations $s_{ij}$
and $\bar s_{ij}$,
the reflections $s_i(\la_i)=-\la_i$.

iii) {\it The $B_N$ root system.}
The positive roots are~:
$$e_i-e_j,\quad e_i+e_j, \quad 1\leq i<j\leq N;
\quad e_i, \quad 1\leq i\leq N\;.$$
The fundamental weights are~:
$$\bar \om_i =e_1+...+e_i, \quad 1\leq i\leq N-1$$ 
$$\bar \om_N =\half(e_1+...e_{N-1}+e_N) $$ and the 
dominant weights are characterised $\la_1\geq...\geq\la_N\geq0$ 
all integers or all half-integers.
The Weyl group is the one 
of $C_N$.

iv) {\it The $BC_N$ root system.} Its positive roots are~: 
$$e_i-e_j,\quad e_i+e_j, \quad 1\leq i<j\leq N;
\quad e_i,\quad 2e_i, \quad 1\leq i\leq N.$$
The Weyl group and the dominant weights
are the ones of $C_N$.

\appendix{A2}{Some Examples of Eigenfunctions}

In this appendix we give as example some eigenvectors of $\CH$ 
at $N=2$. These examples illustrate the "selection rules" of the previous 
appendix, imposed by the different symmetries on the momenta $\la_i$.
The following remarks are valid for any $N$~:

- polynomials labeled by half-integer weights (all $\la_i \in {\bf Z}+1/2$) 
are allowed only for the $B_N$
and $D_N$ cases. 

-for $D_N$, $\la_N\neq 0$, the levels are doubly degenerate, 
$E_{\la_1,\ldots,\la_{N-1}, \la_N}=E_{\la_1,\ldots,\la_{N-1},-\la_N}$.

{\bf $BC_2$}~:

\eqn\polBC{\eqalign{
&\CJ_{1,0}=m_{1,0}+\frac{4c_1}{1+2\be+c_1+2c_2} m_{0,0}\;, \cr
&\CJ_{1,1}=m_{1,1}+\frac{2c_1}{1+c_1+2c_2} m_{1,0}+
\frac{4c_1^2+4\be(1+c_1+2c_2)}
{(1+\be+c_1+2c_2)(1+c_1+2c_2)}m_{0,0}\;,
}}

{\bf $C_2$} $(c_1=0)$~:

\eqn\polC{\eqalign{
&P_{1,0}=m_{1,0}\;, \cr
&P_{1,1}=m_{1,1}+\frac{4c_1^2+4\be(1+2c_2)}
{(1+\be+2c_2)(1+2c_2)}m_{0,0}\;,
}}

{\bf $B_2$} $(c_2=0)$~:

\eqn\polB{\eqalign{
&P_{1/2,1/2}=m_{1/2,1/2}\;, \cr
&P_{1,0}=m_{1,0}+\frac{4c_1}{1+2\be+c_1} m_{0,0}\;, \cr
&P_{3/2,1/2}=m_{3/2,1/2}+\frac{4\be+6c_1}{2+c_1+2\be}m_{1/2,1/2}\;. \cr
}}

{\bf $D_2$} $(c_1=c_2=0)$~:

\eqn\polB{\eqalign{
&P_{1/2,\pm 1/2}=m_{1/2,\pm 1/2}\;, \cr
&P_{1,0}=m_{1,0}\;, \cr
&P_{1,\pm 1}=m_{1,\pm 1}+\frac{2\be}{\be+1} m_{0,0}\;, \cr
&P_{3/2,\pm 1/2}=m_{3/2,\pm 1/2}+\frac{2\be}{\be+1}m_{1/2,\mp 1/2}\;. \cr
}}
The symmetric monomials $m_{\la_1,\la_2}$ associated to each type symmetry 
were defined in \symmon .
 
Unlike in the case of Jack polynomials, there is a $N$ dependendence of 
the coefficients of $m_\la$. 
 
\listrefs

\end